\documentclass[12pt,a4paper,twoside]{article}
\usepackage[english]{babel}
\usepackage[T1]{fontenc}
\usepackage[utf8]{inputenc}
\usepackage{graphicx}
\usepackage{geometry}
\usepackage{amsmath}
\usepackage{amssymb} 
\usepackage{xcolor}
\usepackage{natbib}
\usepackage[legalpaper,bookmarks=true,colorlinks=true,linkcolor=blue,citecolor=blue]{hyperref}
\usepackage{color}
\usepackage{dsfont}
\usepackage{hyperref}
\usepackage[belowskip=-15pt,aboveskip=0pt]{caption}
\usepackage{subcaption}
\usepackage[toc,page]{appendix}
\usepackage{color}
\usepackage{enumitem}

\usepackage{color}
\usepackage{xcolor}
\usepackage{lettrine} 
\usepackage{latexsym}
\usepackage{eurosym}
\usepackage{lmodern}
\usepackage{fancyhdr}
\usepackage{float}
\usepackage{latexsym}
\newcommand{\E}{\mathrm{E}}
\newcommand{\Var}{\mathrm{Var}}
\newcommand{\Cov}{\mathrm{Cov}}
\begin{document}
\pdfoutput=1
\pagenumbering{arabic}
\begin{center}
\textbf{Functional SAC model: With application to spatial econometrics}
\end{center}
Alassane Aw, Department of Mathematics, Assane Seck University of Ziguinchor, Ziguinchor, Senegal.\\
Email: a.aw5720@zig.univ.sn\\
\\
Emmanuel Nicolas Cabral, Department of Mathematics, Assane Seck University of Ziguinchor, Ziguinchor, Senegal.\\
Email: encabral@univ-zig.sn\\
\\
\section*{Abstract}
Spatial autoregressive combined (SAC) model has been widely studied in the literature for the analysis of spatial data in various areas such as geography, economics, demography, regional sciences. This is a linear model with scalar response, scalar explanatory variables and which allows for spatial interactions in the dependent variable and the disturbances. In this work we extend this modeling approach from scalar to functional covariate. The parameters of the model are estimated via the maximum likelihood estimation method. A simulation study is conducted to evaluate the performance of the proposed methodology. As an illustration, the model is used to establish the relationship between unemployment and illiteracy in Senegal.
 \\
 \\
\noindent \textit{Keywords}. Functional linear models, Spatial dependence, Spatial weights.\\
\\
\noindent {\bfseries Mathematics Subject Classification :}  62R10, 62H11, 62P20.\\
\section{Introduction}
Spatially dependent data has become common in many fields such as regional sciences, economics, agriculture and environmental sciences. Consequently, the development of statistical tools dedicated to the modeling of these data has become essential. Spatial statistics is the branch of statistics that deals with this modeling. Various spatial models and methods have been proposed in the literature. Most of them are parametric and deal essentially with non-functional data.\\
Several types of functional linear models for independent data have been developed. They are subdivided into three categories, depending on whether the responses or the regressors, or both, are curves. The common model used in the literature is the functional linear model for scalar response, originally introduced by \cite{Hastie1993}. Estimation and prediction methods for this model has been developed (\cite{Cardot1999}, \cite{Preda2005}, \cite{Ramsay2005}, \cite{Cai2006}). However, in many disciplines of applied sciences, there is a growing need to model correlated functional data. This is the case when samples of functions are observed over a discrete set of time points (temporally correlated functional data) or when these functions are observed over different sites or areas of a region (spatially correlated functional data). In the context of spatially correlated data, some research exists on functional geostatistics (\cite{Dabo}, \cite{Giraldo}, \cite{Caballero}), functional point processes (\cite{Comas2011}, \cite{Comas2013}) and functional areal data (\cite{Zhang2016}, \cite{Ahmed},  \cite{Pineda}, \cite{Aw2019}, \cite{Huang2020}), highlighting the interest in considering spatial functional linear models.\\
This paper deals with spatial functional linear models on lattices. One of the well-known spatial lattice models is the SAC model, which extends regression in time series to spatial data. The methods used to identify and estimate the SAC model are essentially the maximum likelihood (ML) and the generalized method of moments (GMM) (\cite{Anselin1988}, \cite{Kelejian1998}).\\
Specifically, the paper considers an estimation of a spatial functional linear model with a random functional covariate and a real-valued response using spatial autoregression both on the response and the error based on weight matrices. In section \ref{FSACsec2}, we give a short review about the SAC model. Section \ref{FSACsec3} defines its extension to the functional context. The maximum likelihood estimation procedure for the functional SAC model is given in section \ref{FSACsec4}. To check the performance of the ML estimator, numerical results are reported in section \ref{FSACsec5}. We end the article by a conclusion in section \ref{FSACsec6}.
\section{SAC model\label{FSACsec2}}
Let $\{(y_{s}, X_{s}), s\in\mathcal{D}\subset\mathbb{R}^{d}\}$ be a bivariate stochastic process observed over a discrete fixed subset of $\mathcal{D}$ consisting of $s_{1}, s_{2}, \dots, s_{n}$ spatial units and where, for every $s\in\mathcal{D}$, $y_{s}$ and $X_{s}$ are two real-valued random variables. For more flexibility, we will denote the spatial unit $s_{i}$ by $i$. \\
The SAC model is defined (\cite{Anselin1988}) by
\begin{align*}
\begin{split}
y_{i}&=\rho\sum_{j=1}^{n}\omega_{ij}y_{j}+\sum_{j=1}^{p}x_{ij}\beta_{j}+u_{i}\\
u_{i}&=\lambda\sum_{j=1}^{n}m_{ij}u_{j}+\epsilon_{i}, \quad i=1,2,\dots, n,
\end{split}
\end{align*}
or more compactly,
\begin{align}\label{FSACeq1}
\begin{split}
y&= \rho Wy+X\beta+u\\
u&=\lambda Mu+\epsilon,
\end{split}
\end{align}
where 
\begin{itemize}
\item $y$ is an $n\times 1$ vector that collects the dependent variable for each spatial units;\\
\item $W$ and $M$ are $n\times n$ spatial-weighting matrices;\\
\item $X$ is an $n\times p$ matrix of independent variables;\\
\item $\rho$ and $\lambda$ are spatial autoregressive parameters that measure the degree of spatial correlation in the dependent variable $y$ and the disturbance term $u$ respectively;\\
\item $\beta$ is an $p\times 1$ vector of parameters;\\
\item $\epsilon$ is an $n\times 1$ vector of error terms.\\
\end{itemize}
When $\rho=\lambda=0$, the SAC model in (\ref{FSACeq1}) reduces to the classical linear regression model. When $\rho=0$ and $\lambda\neq 0$, it reduces to the SEM model and finaly when $\rho\neq 0$ and $\lambda=0$, the SAC model becomes the spatial lag model.\\
In practice, the spatial-weighting matrices $W$ and $M$ are row-normalized so that $\sum_{j=1}^{n}\omega_{ij}=1$ and $\sum_{j=1}^{n}m_{ij}=1$. In this case, the matrices $I_{n}-\rho W$ and $I_{n}-\lambda M$ are non-singular if $|\rho|<1$ and $|\lambda|<1$. We then obtain the reduced form of the SAC model as
\begin{equation*}
y=(I_{n}-\rho W)^{-1}X\beta+(I_{n}-\rho W)^{-1}(I_{n}-\lambda M)^{-1}\epsilon.
\end{equation*}
The parameters of the SAC model can be obtained by maximum likelihhod estimation method (\cite{Anselin1988}) or generalized spatial two-stage least squares estimation procedure (\cite{Kelejian1998}).
\section{Functional SAC model \label{FSACsec3}}
Consider again $n$ spatial units located in a fixed and countable region $\mathcal{D}\subset\mathbb{R}^{d}, d\in\mathbb{N}^*$. In each spatial unit, we observe a real response variable $y_{i}$  and a functional explanatory variable $X_{i}(t), t\in\mathcal{T}$, where $X_{i}\in L^{2}(\mathcal{T})$ and $\mathcal{T}$ is a compact interval of the real line $\mathbb{R}$. We define the functional SAC model by the following structural equations
\begin{align}\label{FSACeq2}
\begin{split}
y_{i}&=\rho\sum_{j=1}^{n}\omega_{ij}y_{j}+\int_{\mathcal{T}}X_{i}(t)\beta(t)dt+u_{i}\\
u_{i}&=\lambda\sum_{j=1}^{n}m_{ij}u_{j}+\epsilon_{i}, \quad i=1,2,\dots, n,
\end{split}
\end{align}
where $\rho$, $\lambda$, $\omega_{ij}$, $m_{ij}$, $u_{i}$ and $\epsilon_{i}$ are defined as in the SAC model in (\ref{FSACeq1}) and $\beta(.)$ is an unknown functional parameter in $L^{2}(\mathcal{T})$. \\
When $\lambda=0$, the functional SAC model in (\ref{FSACeq2}) reduces to the functional spatial lag model (\cite{Ahmed}) given by 
\begin{equation*}
y_{i}=\rho\sum_{j=1}^{n}\omega_{ij}y_{j}+\int_{\mathcal{T}}X_{i}(t)\beta(t)dt+\epsilon_{i}
, \quad i=1,2,\dots, n.
\end{equation*}
When $\rho=0$, it becomes the functional SEM model (\cite{Pineda}) defined by
\begin{align*}
\begin{split}
y_{i}&=\int_{\mathcal{T}}X_{i}(t)\beta(t)dt+u_{i}\\
u_{i}&=\lambda\sum_{j=1}^{n}m_{ij}u_{j}+\epsilon_{i}, \quad i=1,2,\dots, n.
\end{split}
\end{align*}
When $\rho=\lambda=0$, the functional SAC model reduces to the classical functional linear model (\cite{Cardot1999}) given by
\begin{equation}\label{FSACeq3}
y_{i}=\int_{\mathcal{T}}X_{i}(t)\beta(t)dt+\epsilon_{i}, \quad i=1,2,\dots, n.
\end{equation}
Let $y=\begin{bmatrix} y_{1} & y_{2} & \dots & y_{n}\end{bmatrix}^T$, $X(t)=\begin{bmatrix} X_{1}(t) & X_{2}(t) & \dots & X_{n}(t)\end{bmatrix}^T$, $u=\begin{bmatrix} u_{1} & u_{2} & \dots & u_{n}\end{bmatrix}^T$ and $\epsilon=\begin{bmatrix} \epsilon_{1} & \epsilon_{2} & \dots & \epsilon_{n}\end{bmatrix}^T$, then the matrix form of the functional SAC model is given by
\begin{align}\label{FSACeq4}
\begin{split}
y&= \rho Wy+\int_{\mathcal{T}}X(t)\beta(t)dt+u\\
u&=\lambda Mu+\epsilon,
\end{split}
\end{align}
where $\int_{\mathcal{T}}X(t)\beta(t)dt=\begin{bmatrix}\int_{\mathcal{T}}X_{1}(t)\beta(t)dt & \int_{\mathcal{T}}X_{2}(t)\beta(t)dt & \dots & \int_{\mathcal{T}}X_{n}(t)\beta(t)dt\end{bmatrix}^T$ and where we suppose that $\epsilon\sim N(0, \sigma^2I_{n})$. \\
Consider an orthonormal basis $\{\phi_{j}, j\in \mathbb{N}\}$ of $L^{2}(\mathcal{T})$. We can decompose $X_{i}(t)$ and $\beta(t)$ in this basis as follows
\begin{equation*}
X_{i}(t)=\sum_{j=1}^{\infty}z_{ij}\phi_{j}(t) \quad\mbox{and}\quad \beta(t)=\sum_{j=1}^{\infty}\beta_{j}\phi_{j}(t).
\end{equation*}
The real random variables $z_{ij}$ and the coefficients $\beta_{j}$ are given by
\begin{equation*}
z_{ij}=\int_{\mathcal{T}}X_{i}(t)\phi_{j}(t)dt\quad\mbox{and}\quad\beta_{j}=\int_{\mathcal{T}}\beta(t)\phi_{j}(t)dt.
\end{equation*}
From this decomposition, it follows that
\begin{equation}\label{FSACeq5}
\int_{\mathcal{T}}X_{i}(t)\beta(t)dt=\sum_{j=1}^{\infty}z_{ij}\beta_{j}.
\end{equation}
Denoting $z_{j}=\begin{bmatrix} z_{1j} & z_{2j} & \dots & z_{nj}\end{bmatrix}^T$, we can can deduce from (\ref{FSACeq5}) that
\begin{equation}\label{FSACeq6}
\int_{\mathcal{T}}X(t)\beta(t)dt=\begin{bmatrix}
\int_{\mathcal{T}}X_{1}(t)\beta(t)dt\\
\int_{\mathcal{T}}X_{2}(t)\beta(t)dt\\
\vdots\\
\int_{\mathcal{T}}X_{n}(t)\beta(t)dt\\
\end{bmatrix}=\sum_{j=1}^{\infty}z_{j}\beta_{j}.
\end{equation}
The estimation of the functional parameter $\beta(t)$ requires a regularization procedure. Let us project $X(t)$ and $\beta(t)$ onto a finite dimensional space spanned by $K$ basis functions $\phi_{1}(t),\phi_{2}(t), \dots, \phi_{K}(t)$. We can therefore rewrite (\ref{FSACeq6}) as follows
\begin{equation}\label{FSACeq7}
\int_{\mathcal{T}}X(t)\beta(t)dt=\sum_{j=1}^{K}z_{j}\beta_{j}
\end{equation}
The truncated equation (\ref{FSACeq7}) can be obtained by using the Fourier basis, the functional principal components basis, the partial least squares basis, etc. In the following, we use the functional partial least squares (FPLS)  basis functions. To form the FPLS basis functions, we first neglect the autoregressive terms by considering model (\ref{FSACeq3}), which means building the basis without considering any spatial correlation (see \cite{Huang2020}). We then use the iterative process introduced by \cite{Preda2005} to obtain the basis. \\
The main steps for constructing a number $K$ of FPLS basis functions are given below. \\
Step 1. Begin from $k=1$ and set $X_{k}(t)=X(t)$, $y_{k}=y$.\\
Step 2. Find a square integrable weight function $\omega_{k}(t)$ that maximizes the following covariance $\underset{||\omega_{k}(t)=1||}\Cov\left(y_{k}, \int_{\mathcal{T}}X_{k}(t)\omega_{k}(t)dt\right)$. We obtain $\omega_{k}(t)=\frac{\E(y_{k}X_{k}(t))}{||\E(y_{k}X_{k}(t))||}$, where $||.||$ denotes the usual norm of $L^{2}(\mathcal{T})$, i.e, $||\omega_{k}(t)||=\sqrt{\int_{\mathcal{T}}\omega_{k}^{2}(t)dt}$.\\ 
Step 3. Define $z_{k}=\int_{\mathcal{T}}X_{k}(t)\omega_{k}(t)dt$ and perform the regressions $X_{k}(t)=a_{k}(t)z_{k}+\epsilon_{k}^{X}(t)$ and $y_{k}=b_{k}z_{k}+\epsilon_{k}^{y}$. We obtain $a_{k}(t)=\frac{\E(X_{k}(t)z_{k})}{||z_{k}^2||}$ and $b_{k}=\frac{\E(y_{k}z_{k})}{||z_{k}^2||}$.\\
Step 4. Stop when k=K. Otherwise, take $X_{k+1}(t)=\epsilon_{k}^{X}(t)$ and $y_{k+1}=\epsilon_{k}^{y}$  and go back to step 2. \\ 
The weight functions $\omega_{1}(t), \omega_{2}(t), \dots, \omega_{K}(t)$ produced by the above iterative procedure are the functional partial least squares basis functions. In practice, we take the empirical versions of the quantities involved in the iterative procedure described above. We then take $\phi_{k}(t)=\hat{\omega}_{k}(t)$ and $z_{k}=\int_{\mathcal{T}}X_{k}(t)\phi_{k}(t))dt$ for $k=1, 2, \dots, K$. \\
\\
Let $\textbf{Z}_{K}=\begin{bmatrix} z_{1} & z_{2} & \dots & z_{K}\end{bmatrix}^T$ and $\mathbf{\beta_{K}}=\begin{bmatrix} \beta_{1} & \beta_{2} & \dots & \beta_{K}\end{bmatrix}^T$, then the truncated equation (\ref{FSACeq7}) is written as
\begin{equation}\label{FSACeq8}
\int_{\mathcal{T}}X(t)\beta(t)dt=\textbf{Z}_{K}\mathbf{\beta_{K}}
\end{equation} 
\section{Maximum likelihood estimation\label{FSACsec4}}
From (\ref{FSACeq4}), we have
\begin{align*}
\begin{split}
(I_{n}-\rho W)y&= \int_{\mathcal{T}}X(t)\beta(t)dt+u\\
(I_{n}-\lambda M)u &=\epsilon
\end{split}
\end{align*}
Noting $A=I_{n}-\rho W$ and $B=I_{n}-\lambda M$ and solving for $\epsilon$, we obtain
\begin{equation}\label{FSACeq9}
\epsilon=B\left(Ay-\int_{\mathcal{T}}X(t)\beta(t)dt\right).
\end{equation}
Using the transformation theorem, the probability density function of $y$ is given by
\begin{equation*}
f(y|\beta(t), \sigma^2, \rho, \lambda, X(t))=f(\epsilon|\beta(t), \sigma^2, \rho, \lambda, X(t))\left|\frac{\partial \epsilon}{\partial y}\right|,
\end{equation*}
where $\frac{\partial \epsilon}{\partial y}$ is the Jacobian matrix. Since $\epsilon\sim N(0, \sigma^2 I_{n})$, the probability density of $\epsilon$ is expressed as
\begin{equation*}
f(\epsilon|\beta(t), \sigma^2, \rho, \lambda, X(t))=(2\pi)^{-\frac{n}{2}}(\sigma^2)^{-\frac{n}{2}}\exp\left[-\frac{1}{2\sigma^2}\epsilon^T\epsilon\right],
\end{equation*}
where $\epsilon^T\epsilon=\left(Ay-\int_{\mathcal{T}}X(t)\beta(t)dt\right)^T B^TB\left(Ay-\int_{\mathcal{T}}X(t)\beta(t)dt\right)$ and $B^TB=\Omega(\lambda)=(I_{n}-\lambda M)^T(I_{n}-\lambda M)$.\\
The Jacobian matrix is derived, from (\ref{FSACeq9}), by
\begin{equation*}
\left|\frac{\partial \epsilon}{\partial y}\right|=\left|\frac{\partial B\left(Ay-\int_{\mathcal{T}}X(t)\beta(t)dt\right)}{\partial y}\right|=\left|BA\right|=|B||A|.
\end{equation*}
Consequently, the likelihood function is given by
\begin{equation*}
L=(2\pi)^{-\frac{n}{2}}(\sigma^2)^{-\frac{n}{2}}\exp\left[-\frac{1}{2\sigma^2}\left(Ay-\int_{\mathcal{T}}X(t)\beta(t)dt\right)^T \Omega(\lambda)\left(Ay-\int_{\mathcal{T}}X(t)\beta(t)dt\right)\right]|B||A|.
\end{equation*}
The log-likelihood function $l=\log(L)$ has the following expression\\
\\
$l=-\frac{n}{2}\log(2\pi)-\frac{n}{2}\log(\sigma^2)-\frac{1}{2\sigma^2}\left(Ay-\int_{\mathcal{T}}X(t)\beta(t)dt\right)^T \Omega(\lambda)\left(Ay-\int_{\mathcal{T}}X(t)\beta(t)dt\right)+\log(|A|)$\\
$+\log(|B|).
$
\\
Using the truncation equation (\ref{FSACeq8}), we can then define the truncated log-likelihood function $l_{K}$ as follows\\
\begin{equation}\label{FSACeq10}
l_{K}=-\frac{n}{2}\log(2\pi)-\frac{n}{2}\log(\sigma^2)-\frac{1}{2\sigma^2}\left(Ay-\textbf{Z}_{K}\mathbf{\beta_{K}}\right)^T \Omega(\lambda)\left(Ay-\textbf{Z}_{K}\mathbf{\beta_{K}}\right)+\log(|A|)+\log(|B|).
\end{equation}
Taking the derivative of $l_{K}$ with respect to $\mathbf{\beta_{K}}$ yields
\begin{eqnarray}
\frac{\partial l_{K}}{\partial\mathbf{\beta_{K}}} 
                                                 &=& -\frac{1}{2\sigma^2}\frac{\partial}{\partial\mathbf{\beta_{K}}}\left(Ay-\textbf{Z}_{K}\mathbf{\beta_{K}}\right)^T \Omega(\lambda)\left(Ay-\textbf{Z}_{K}\mathbf{\beta_{K}}\right)\nonumber\\
                                                 &=& -\frac{1}{2\sigma^2}\left[2\frac{\partial \left(Ay-\textbf{Z}_{K}\mathbf{\beta_{K}}\right)^T}{\partial\mathbf{\beta_{K}}}\Omega(\lambda)\left(Ay-\textbf{Z}_{K}\mathbf{\beta_{K}}\right)\right]\nonumber\\
                                                 &=& -\frac{1}{\sigma^2}\frac{\partial (-\textbf{Z}_{K}\mathbf{\beta_{K}})^T}{\partial\mathbf{\beta_{K}}}\Omega(\lambda)\left(Ay-\textbf{Z}_{K}\mathbf{\beta_{K}}\right)\nonumber\\
                                                 &=& \frac{1}{\sigma^2}\textbf{Z}_{K}^T\Omega(\lambda)\left(Ay-\textbf{Z}_{K}\mathbf{\beta_{K}}\right)\nonumber.
\end{eqnarray}
The ML estimator of $\mathbf{\beta_{K}}$ is obtained by solving the equation $\frac{\partial l_{K}}{\partial\mathbf{\beta_{K}}}=0$, which gives
\begin{equation}\label{FSACeq11}
\hat{\beta}_{ML}(\rho, \lambda)=[\textbf{Z}_{K}^T\Omega(\lambda)\textbf{Z}_{K}]^{-1}\textbf{Z}_{K}^T\Omega(\lambda)Ay.
\end{equation}
The derivative of $l_{K}$ with respect to $\sigma^2$ is given by
\begin{eqnarray}
\frac{\partial l_{K}}{\partial \sigma^2} 
                                                 &=& -\frac{n}{2}\frac{\partial \log(\sigma^2)}{\partial \sigma^2}-\frac{1}{2}\frac{\partial}{\partial \sigma^2}\left(Ay-\textbf{Z}_{K}\mathbf{\beta_{K}}\right)^T \Omega(\lambda)\left(Ay-\textbf{Z}_{K}\mathbf{\beta_{K}}\right)\nonumber\\
                                                 &=& -\frac{n}{2\sigma^2}+\frac{1}{2\sigma^4}\left(Ay-\textbf{Z}_{K}\mathbf{\beta_{K}}\right)^T \Omega(\lambda)\left(Ay-\textbf{Z}_{K}\mathbf{\beta_{K}}\right)\nonumber\\
                                                 &=& \frac{-n\sigma^2+\left(Ay-\textbf{Z}_{K}\mathbf{\beta_{K}}\right)^T \Omega(\lambda)\left(Ay-\textbf{Z}_{K}\mathbf{\beta_{K}}\right)}{2\sigma^4}\nonumber.
\end{eqnarray}
Solving $\frac{\partial l_{K}}{\partial \sigma^2}=0$, we find the ML estimator of $\sigma^2$ given by
\begin{equation}\label{FSACeq12}
\hat{\sigma}_{ML}^2(\rho, \lambda)=\frac{\left(Ay-\textbf{Z}_{K}\hat{\beta}_{ML}(\rho, \lambda)\right)^T \Omega(\lambda)\left(Ay-\textbf{Z}_{K}\hat{\beta}_{ML}(\rho, \lambda)\right)}{n}.
\end{equation}
Then, plugging (\ref{FSACeq11}) and (\ref{FSACeq12}) into the truncated log-likelihood (\ref{FSACeq10}), we obtain the truncated concentrated log-likelihood function given below
\begin{equation}\label{FSACeq13}
l_{c}=-\frac{n}{2}-\frac{n}{2}\log(2\pi)-\frac{n}{2}\log(\hat{\sigma}_{ML}^2(\rho, \lambda))+\log(|I_{n}-\rho W|)+\log(|I_{n}-\lambda M|).
\end{equation}
The ML estimators of the autoregressive parameters $\rho$ and $\lambda$ are obtained by maximizing the function $l_{c}$, which is highly nonlinear. Numerical optimization methods are used to solve (\ref{FSACeq13}). After obtaining $\hat{\rho}$ and $\hat{\lambda}$, the estimators of $\mathbf{\beta_{K}}$ and $\sigma^2$ are recalculated by setting $\hat{\beta}_{ML}=\hat{\beta}_{ML}(\hat{\rho}, \hat{\lambda})$ and $\hat{\sigma}^2_{ML}=\hat{\sigma}_{ML}^2(\hat{\rho}, \hat{\lambda})$.\\
The estimator of the functional parameter $\beta(t)$ is given by
\begin{equation}\label{FSACeq14}
\hat{\beta}_{ML}(t)=\hat{\beta}_{ML}^T\Phi(t),
\end{equation}
where $\Phi(t)=(\phi_{1}(t), \dots, \phi_{K}(t))^T$ is the vector of the $K$ functional partial least squares basis functions.\\
Since
\begin{align*}
\begin{split}
Ay&= \int_{\mathcal{T}}X(t)\beta(t)dt+u\\
Bu &=\epsilon,
\end{split}
\end{align*}
we have 
\begin{equation}\label{FSACeq15}
y=A^{-1}\int_{\mathcal{T}}X(t)\beta(t)dt+A^{-1}B^{-1}\epsilon=A^{-1}\textbf{Z}_{K}\mathbf{\beta_{K}}+A^{-1}B^{-1}\epsilon.
\end{equation}
Using (\ref{FSACeq15}), we obtain
\begin{eqnarray}
\label{FSACeq16}
\E(y) &=& A^{-1}\textbf{Z}_{K}\mathbf{\beta_{K}} \\
\label{FSACeq17}
\Var(y) &=& \sigma^{2}A^{-1}\Omega^{-1}(\lambda)(A^{T})^{-1}.
\end{eqnarray}
Now, let us assume that $\rho$ and $\lambda$ are known and $K$ is fixed. Set $\Sigma_{K}=\textbf{Z}_{K}^T\Omega(\lambda)\textbf{Z}_{K}$. We can then write
\begin{equation}\label{FSACeq18}
\hat{\beta}_{ML}=\Sigma_{K}^{-1}\textbf{Z}_{K}^T\Omega(\lambda)Ay.
\end{equation}
We deduce, from (\ref{FSACeq16}) and (\ref{FSACeq18}), that
\begin{eqnarray}
\E(\hat{\beta}_{ML}) &=& \Sigma_{K}^{-1}\textbf{Z}_{K}^T\Omega(\lambda)A\E(y)\nonumber\\
                     &=& \Sigma_{K}^{-1}\textbf{Z}_{K}^T\Omega(\lambda)AA^{-1}\textbf{Z}_{K}\mathbf{\beta_{K}}\nonumber\\
                     &=& \Sigma_{K}^{-1}\textbf{Z}_{K}^T\Omega(\lambda)\textbf{Z}_{K}\mathbf{\beta_{K}}\nonumber\\
                     &=& \Sigma_{K}^{-1}\Sigma_{K}\mathbf{\beta_{K}}\nonumber\\
                     &=& \mathbf{\beta_{K}}\nonumber.
\end{eqnarray}
Hence, $\hat{\beta}_{ML}$ is an unbiased estimator of $\mathbf{\beta_{K}}$. We can also deduce, from (\ref{FSACeq17}) and (\ref{FSACeq18}), that
\begin{eqnarray}
\Var(\hat{\beta}_{ML}) &=& \Sigma_{K}^{-1}\textbf{Z}_{K}^T\Omega(\lambda)A\Var(y)A^{T}\Omega(\lambda)\textbf{Z}_{K}\Sigma_{K}^{-1}\nonumber\\
                       &=& \sigma^{2}\Sigma_{K}^{-1}\textbf{Z}_{K}^T\Omega(\lambda)AA^{-1}\Omega^{-1}(\lambda)(A^{T})^{-1}A^{T}\Omega(\lambda)\textbf{Z}_{K}\Sigma_{K}^{-1}\nonumber\\
                       &=& \sigma^{2}\Sigma_{K}^{-1}\textbf{Z}_{K}^T\Omega(\lambda)\textbf{Z}_{K}\Sigma_{K}^{-1}\nonumber\\
                       &=& \sigma^{2}\Sigma_{K}^{-1}\Sigma_{K}\Sigma_{K}^{-1}\nonumber\\
                       &=& \sigma^{2}\Sigma_{K}^{-1}\nonumber.
\end{eqnarray}
Therefore, a confidence band of $100(1-\alpha)\%$ of the functional parameter $\beta(t)$ is given by
\begin{equation}\label{FSACeq19}
\left[\hat{\beta}_{ML}(t)-z_{1-\frac{\alpha}{2}}\hat{\sigma}_{ML}\sqrt{\Phi^{T}(t)\Sigma_{K}^{-1}\Phi(t)},\hat{\beta}_{ML}(t)+z_{1-\frac{\alpha}{2}}\hat{\sigma}_{ML}\sqrt{\Phi^{T}(t)\Sigma_{K}^{-1}\Phi(t)}\right]. 
\end{equation}
In practice, $\rho$ and $\lambda$ are unknown. We replace them by their ML estimators $\hat{\rho}$ and $\hat{\lambda}$ respectively. We can then use $\hat{\lambda}$ to estimate $\Sigma_{K}$ and as a result obtain the confidence band (\ref{FSACeq19}).
\section{Numerical results\label{FSACsec5}} 
In this section, we study the performance of the proposed model based on numerical results. We use the partial least squares functions obtained from the iterative procedure described in section \ref{FSACsec3} to construct the expansion basis. For the choice of the optimal number $K$ of functions included in the truncation strategy, we use the Bayesian information criterion to select it.
\subsection{Simulation results}
To carry out the simulations, we use the spatial layout of 121 communes of Senegal represented in figure \ref{figure1} below. 
\begin{figure}[H] 
\centering
\includegraphics[width=16cm,height=8cm]{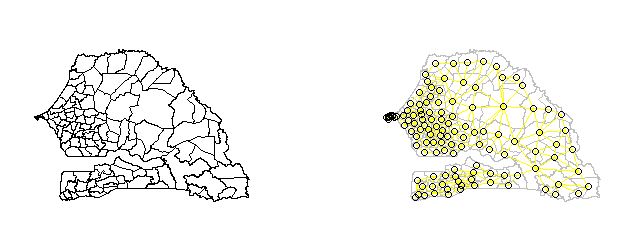}
\caption{Spatial layout of communes of Senegal (left panel) and centroids of these communes (right
panel).} 
\label{figure1}
\end{figure}
\noindent We consider several scenarios of simulations according to the values of the autoregressive parameters $\rho$ and $\lambda$. Specifically, we use the following steps:
\begin{enumerate}
\item A spatial $121\times 121$ row-standardized weight matrix W is calculated using the first order contiguity relations of the 121 communes in right panel of figure \ref{figure1} above;
\item In each commune, we simulate $X_{i}(t), i=1,2,\dots, 121$ as a Brownian motion. All the curves $X_{i}(t)$ are discretized on the same grid generated from 101 equispaced points $t\in [0, 1]$.
\item The functional parameter is defined by $\beta(t)=t\sin(\pi t)^2$;
\item We generate a $121\times 1$ Gaussian vector $\epsilon\sim N(0, I_{n})$;
\item We calculate $y=(I_{n}-\rho W)^{-1}\int_{\mathcal{T}}X(t)\beta(t)dt+(I_{n}-\rho W)^{-1}(I_{n}-\lambda W)^{-1}\epsilon$ by considering the pairs $(\rho, \lambda)=(0.1, 0.9), (0.3, 0.7), (0.5, 0.5), (0.7, 0.3), (0.9, 0.1)$;
\item The parameters $\beta(t)$, $\sigma^2$, $\rho$ and $\lambda$ are estimated using the maximum likelihood estimation procedure described in section \ref{FSACsec4};
\item The steps 4, 5 and 6 are repeated 500 times. In each case, we calculate $\hat{\beta_{j}}(t)$, $\hat{\sigma}^2_{j}$, $\hat{\rho}_{j}$, $\hat{\lambda}_{j}$ and the integrated squared error of $\hat{\beta_{j}}(t)$ defined by $ISE=\int_{0}^{1}(\hat{\beta_{j}}(t)-\beta(t))^2dt$;
\item Finaly, we calculate $\hat{\sigma}^2=\frac{1}{500}\sum_{j=1}^{500}\hat{\sigma}^2_{j}$, $\hat{\rho}=\frac{1}{500}\sum_{j=1}^{500}\hat{\rho}_{j}$, $\hat{\lambda}=\frac{1}{500}\sum_{j=1}^{500}\hat{\lambda}_{j}$ and the mean of integrated squared error (MISE).
\end{enumerate}
The results of the procedure described above are shown in table \ref{tableau1} below. 
\begin{table}[!htbp]
\centering
\caption{Summary statistics of estimations for the spatial autoregressive parameters  $\rho$ and $\lambda$, the variance $\sigma^2$ and the mean of integrated squared error (MISE).}
\vspace{0.5cm}
 \begin{tabular}{||c c c c c c||} 
 \hline
 $\rho$ & $\lambda$ & $\hat{\rho}$ & $\hat{\lambda}$ & $\hat{\sigma}^2$ & MISE\\ [0.5ex] 
 \hline
 0.1 & 0.9 & 0.08 & 0.87& 0.99 & 0.17 \\ 
 0.3 & 0.7 & 0.31 & 0.68& 0.94 & 0.14 \\
 0.5 & 0.5 & 0.51 & 0.49 & 0.94 & 0.16 \\
 0.7 & 0.3 & 0.68 & 0.27 & 0.93 & 0.15\\
 0.9 & 0.1 & 0.88 & 0.09 & 0.93 & 0.13\\ [1ex] 
 \hline
 \end{tabular}
\label{tableau1}
\end{table}
\noindent \\The results obtained in table \ref{tableau1} indicate that the maximum likelihood estimation procedure allows to accurately estimate the spatial autoregressive parameters $\rho$ and $\lambda$. We can also notice, with a $K$ small number of partial least squares basis functions , are obtained estimations slightly biased of $\sigma^2$ and $\beta(t)$. We can then conclude from a practical point of view that the methodology proposed has a good performance.\\
We give in figure \ref{figure2} below the estimations of the functional parameter $\beta(t)$ when $\rho=\lambda=0.5$ and $\sigma^2=1$, which confirms the good performance of the methodology proposed.
\begin{figure}[H]
\centering
\includegraphics[width=16cm,height=8cm]{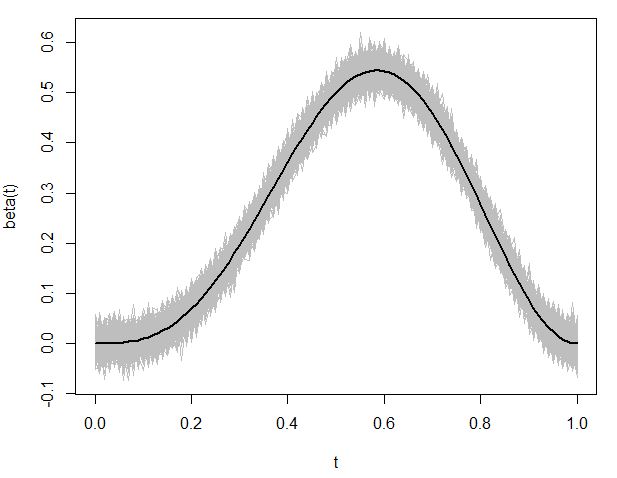} 
\caption{Functional parameter $\beta(t)$ (black line) and 500 estimations of this one (gray lines) obtained with $\rho=\lambda=0.5$ and $\sigma^2=1$.} 
\label{figure2}
\end{figure}
\subsection{Relationship Between Unemployment and Illiteracy\label{FSACsubsec2}}
\begin{figure}[H]
\centering
\includegraphics[width=16cm,height=8cm]{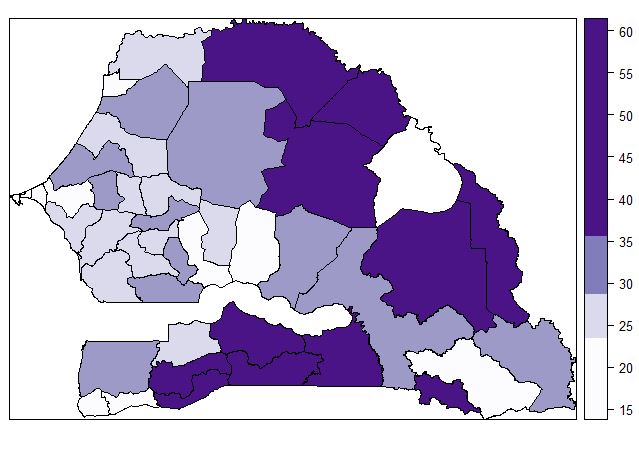}
\caption{Choropleth map of unemployment in Senegal.} 
\label{figure3}
\end{figure}

We now apply the proposed methodology to real data. The data relate to unemployment rates and illiteracy rates observed in the various departments of Senegal. More precisely, in each of the 45 departments of Senegal, we observe the unemployment rate in the first quarter of 2019 as well as the illiteracy rates ranging from the second quarter of 2016 to the first quarter of 2019. The data come from the National Agency of Statistics and Demography of Senegal (ANSD, by its acronym in French). Our goal is to establish the relationship between these variables as a contribution to explain the regional variation of unemployment in Senegal. The unemployment rate is taken to be our scalar response and the illiteracy rate as the functional covariate. As a first step, we proceed to the description of the behavior of the variables of interest.\\
We show in figure \ref{figure3} below the choropleth map of the unemployment rates in the 45 departments of Senegal. It is easy to see in this figure that the spatial distribution of the unemployment rate is not due to mere chance. Indeed, nearby departments tend to have similar unemployment rates. This phenomenon is known in the literature as spatial autocorrelation. The presence of spatial autocorrelation in the distribution of the unemployment rate was confirmed by the Moran test. 
\noindent 
Since the illiteracy rates are recorded discretely, we smooth them using a 7 B-spline basis functions in order to obtain the curves as functional covariates. This number of basis is selected by using cross-validation criterion (\cite{Ramsay2005}). \\
Figure \ref{figure4} below shows the curves of the illiteracy rates observed in the 45 departments of Senegal over the observation period. There is a decrease in the illiteracy curves over the entire observation period. This shows the great efforts made
by the Government of Senegal to reduce the illiteracy rate by formulating targeted literacy
programs.
\begin{figure}[H]
\centering
\includegraphics[width=16cm,height=8cm]{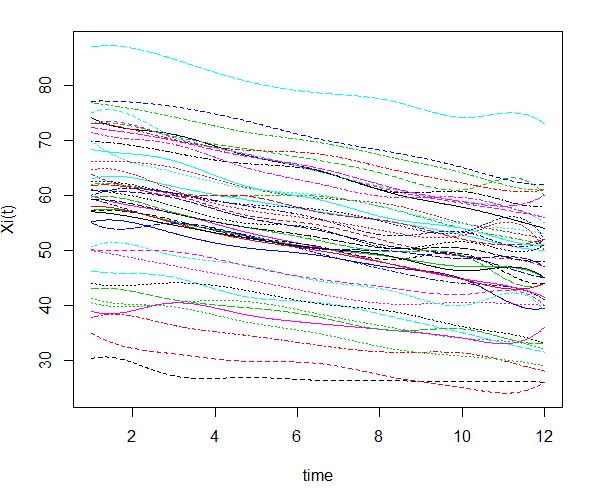}
\caption{Curves of illiteracy observed in the 45 departments of Senegal.} 
\label{figure4}
\end{figure}
\noindent We use the FSAC model to establish the relationship between unemployment and illiteracy. We focus on the estimation of the functional parameter $\beta(t)$. This can be interpreted as the impact of illiteracy on the unemployment of the first quarter of 2019. We give in figure \ref{figure5} below the estimation of $\beta(t)$ and its 95$\%$ confidence bands.
\begin{figure}[H]
\centering
\includegraphics[width=16cm,height=8cm]{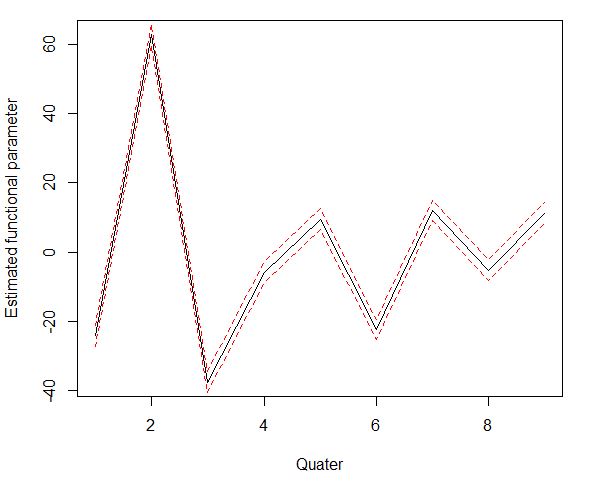}
\caption{Estimation of $\beta(t)$ (black line) and 95$\%$ confidence bands (red lines).} 
\label{figure5}
\end{figure}
\noindent According to the estimation in figure 5, we conclude that illiteracy has a varied impact on unemployment during the study period. The impact is greater in the first 4 quarters. Then, it decreases as we advance in the other quarters.
\section{Conclusion\label{FSACsec6}}
In this article, we defined the functional SAC model for areal data and developed a methodology to make inference. The proposed model can be seen as an extension of the real-valued SAC model to a functional model. The proposed maximum likelihood estimation approach based on a truncation technique is particularly well adapted to spatial regression estimation for functional data in the presence of spatial
dependence. The application of the methodology to real data shows that illiteracy has a real impact on unemployment in Senegal. This work offers interesting perspectives for investigation. For example, an adaptation of this method to issues using different covariates (functional and non-functional) could be developed. Also, the application of this methodology to other types of data and other areas of activity could be considered.

\end{document}